# Solutions for 400 Gbit/s Inter Data Center WDM Transmission

Annika Dochhan[(1)], Nicklas Eiselt[(1,2)], Helmut Griesser[(3)], Michael Eiselt[(1)], Juan José Vegas Olmos[(2)], Idelfonso Tafur Monroy[(2)], Jörg-Peter Elbers [(3)]

[(1)] ADVA Optical Networking SE, Maerzenquelle 1-3, 98617 Meiningen, Germany, dochhan@advaoptical.com
[(2)] Technical University of Denmark (DTU), Dep. of Photonics Engineering, Ørsteds Plads, Build. 343, DK-2800
[(3)] ADVA Optical Networking SE, Fraunhoferstr. 9a, 82152 Martinsried, Germany

**Abstract** *We review some currently discussed solutions for 400 Gbit/s inter-data center WDM transmission for up to 100 km. We focus on direct detected solutions, namely PAM4 and DMT, and present two WDM systems based on these formats.*

**Introduction**

Recent estimations of the future data center traffic forecast a growth of 25% per year[1]. Special solutions need to be developed for the traffic between data centers, bridging a distance of up to 80 km. This application amounts to 7% of all data traffic. In contrast to short-reach links within the data center (intra-data center), where standardization efforts are currently ongoing[2], for the links between data centers (inter-data-center), no standards are currently being developed. To keep implementation cost and power consumption low, the re-use of intra-data center technology (with some add-ons) appears as a reasonable solution. Therefore, promising modulation formats for this purpose are directly detected (DD) formats like four-level pulse amplitude modulation (PAM4) and discrete multi-tone transmission (DMT). Carrier-less amplitude modulation (CAP) in various flavours was also proposed for this purpose[3,4], but since digital signal processing (DSP) technology is already available for PAM4 and DMT[5-7], in this paper we concentrate on these two formats.

Instead of up-scaling from short-reach technology to inter-data center connections, a down-scaling from long-haul solutions using coherent reception and quadrature amplitude modulation (QAM) formats with polarization division multiplexing (PDM) would be an alternative. PDM-64QAM appears as an adequate solution, mainly due to its much higher spectral efficiency than the DD solutions[8]. However, in the very near future, the cost, footprint and power consumption requirements seem to be better matched using DD formats, which is the reason for our focus on them.

**Requirements for data center interconnects**

Besides low cost, low power consumption and small footprint, the following requirements have to be met: Link capacities should cover several Tb/s, in multiples of 400 Gbit/s. The reach should cover distances up to 100 km, while many links might be also shorter, suggesting the use of flexible systems as proposed below. Finally, the transmission should use the C-band to enable the application of Erbium doped fiber amplifiers (EDFAs). Dense wavelength division multiplexing (DWDM) should be applied to enable the desired rates. As mainly standard single mode fiber (SSMF) is deployed between data centers, the receiver needs to have some tolerance towards chromatic dispersion (CD). If compensation of CD can be completely omitted is a matter of the chosen modulation format, channel data rate and DSP complexity as will be discussed below. To enable 400G transmission with PAM4 or DMT, either eight channels with 50G per channel can be used or four channels with 100G are possible. A flexible intermediate solution is also proposed.

**Solutions for 400G**

100 Gbit/s transmission enabled by DMT was first shown by Yan[9], demonstrating a reach of 10 km using C-band lasers. Later experiments showed 10 and 30 km reach with four WDM lanes[10,11] (more than 448 Gbit/s), and 100 Gbit/s over 80 km SSMF, enabled by a semiconductor optical amplifier (SOA)[12] in the O-band. However, DWDM transmission requires the use of EDFAs and thus C-band transmission is mandatory. Several low-cost modulator techniques have been demonstrated so far, including directly modulated lasers (DMLs)[9], electro absorption modulators (EMLs)[10] and integrated arrays of Silicon Photonic Mach-Zehnder modulators (SiPh-MZMs)[13]. For short reach applications up to 4 km, even the use of vertical cavity surface emitting lasers (VCSELs)[14] and Si-Ph ring modulators[15] has been demonstrated. To enable 100 Gbit/s transmission with up to 80 km reach, several single-side band (SSB) or vestigial side band (VSB) modulation techniques have been proposed, among which simple asymmetrical filtering is the least complex one[13,16,17]. More sophisticated techniques include an optical IQ modulator to enable real SSB modulation or at least a dual-drive MZM, and thus two independent electrical driving signals from the digital-to-analog converter[18-20]. Only one of the proposed solutions[19] might enable a 50-GHz channel grid, since a 30 GHz optical receiver filter was included. Other solutions used wider bandwidths or channel grids[13]. If a 50-GHz-channel grid is used, lower data rate signals should be considered, e.g. up to 74.7 Gbit/s[16], or around 50 Gbit/s as demonstrated with 16QAM-DMT[21] in a 50-GHz channel grid and 128QAM- and 512QAM-DMT in a 25-GHz-grid[22] with real-time reception. To improve the performance, several advanced DSP techniques have been included, such as non-linear Volterra equalization[17,19,20], subcarrier-interference cancellation[18], trellis coded modulation[19,20], DFT spreading[21,22] and pre-equalization[22]. The results indicate that in future applications, 400 Gbit/s transmission in four 50-GHz-spaced channels will be possible.

Multilevel-signalling like PAM4 was already early proposed for optical communication at low data rates[24] but attracted much attention during the last years especially for short reach applications with few kilometres reach. The performance is strongly dependent on the applied pre- and post-equalization, especially, if longer reaches than 2 km are targeted or low-cost, low bandwidth components are used. In contrast to DMT, PAM4 is very sensitive to chromatic dispersion, which might be the reason why most applications target at a transmission in the O-band[5,6]. The longest reach in the C-band for 56 Gbit/s PAM4 without optical CD compensation was 26.4 km, enabled by non-linear filtering, decision feedback equalization (DFE) and maximum likelihood sequence estimation (MLSE)[25]. Also enabled by MLSE, 100 Gbit/s transmission over 5 km without and 80 km with optical CD compensation are possible[26]. 100 Gbit/s PAM4 up to 4 km were achieved using a high-speed selector power digital-to-analog converter (DAC)[27]. With this device, 5 km transmission of 75 Gbit/s was shown for 2D-coded PAM4, and PAM8 was transmitted over 2 km[28]. Even 5-km 100-Gbit/s transmission and 10-km 50-Gbit/s transmission were shown with Duobinary-PAM4[29,30]. Very high spectral efficiency (25-GHz channel grid) and 20 km reach was achieved with quasi SSB Nyquist PAM4[31]. These publications indicate that without optical CD compensation PAM4 transmission over the desired reach is not achievable so far. Only some dispersion tolerance can be gained using the proposed solutions. Therefore, in our proposed system, we will also include a dispersion compensating fiber (DCF) to enable 100 km transmission.

**DMT: Flexible DWDM 400G system**
Due to the flexibility of DMT enabled by adaptive bit loading, the data rate of the system can be changed easily without the need of variable clock generators and other component changes. We proposed a flexible 400G transmission system with up to eight DWDM channels, depending on the desired reach[16,23]. Five DWDM channels with a rate of 89.6 Gbit/s could be transmitted over 40 km, whereas six channels with 74.7 Gbit/s are needed for 80 km. Even longer reaches, such as 160 and 240 km were achieved for seven and eight channels with 64 and 56 Gbit/s respectively. Fig. 1 shows a schematic plot of the proposed system. The reach was enabled by VSB filtering[16]. It also shows the bit error ratios (BER) after transmission for each channel after the maximum reach. Details can be found in our previous publications[16,23].

**PAM4: Real-time DWDM 400G system**
The system setup for the real-time 400G system is shown in Fig. 3. Real-time generation and decoding of two 25.78125-GBaud PAM4 signals was achieved by the Inphi IN015025-CA0 PAM4 transceiver PHY. An eight-channel DWDM system with 50-GHz channel grid was created by interleaving two four-channel, 100-GHz grid spaced and simultaneously modulated optical signals. Our recent publications describe further details of the system and also single channel properties of the PAM4 signal[32,33]. With only simple 3-tap pre-equalization and a level adjusting function at the transmit side and feed forward (FFE)-DFE equalization at the receiver, all eight channels stayed well below the KR4 forward error correction (FEC) limit of $5.2 \cdot 10^{-5}$, allowing error free transmission. Fig. 2 shows the achieved performance together with the CD tolerance of single channel 50 Gbit/s PAM4.

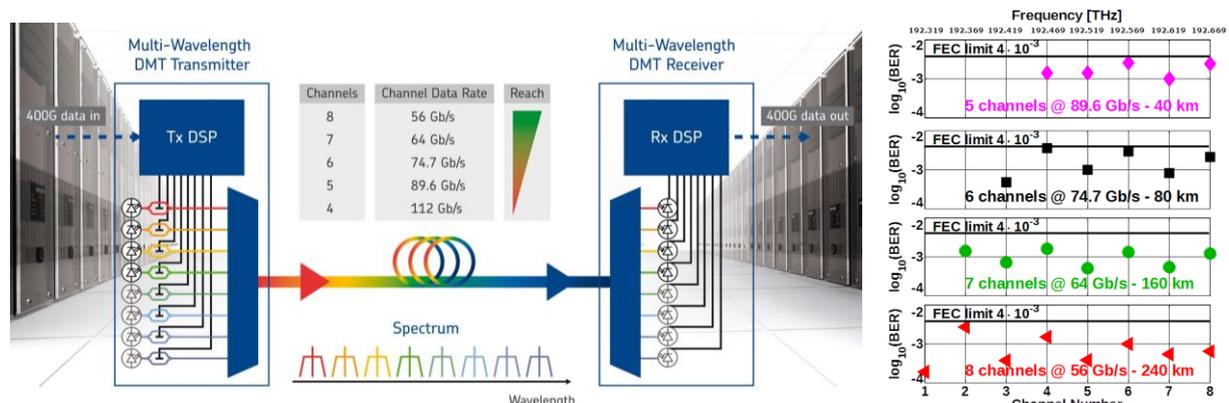

**Fig. 1:** Schematic of the flexible 400G DMT system (left), BER for all channel counts, data rates and reaches (right).

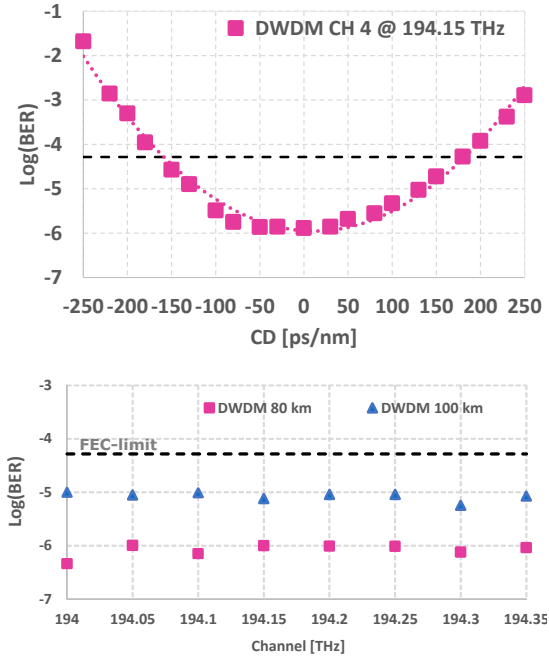

**Fig. 2:** Tolerance of 50 Gbit/s PAM4 to residual dispersion (upper), DWDM transmission results after 80 and 100 km.

## Conclusions

We reviewed recently proposed solutions for 400G inter-data center interconnects. While PDM-64-QAM offers highest spectral efficiency and flexibility in terms of dispersion and other distortion compensation, its advent as commercial product that can outperform directly detected solutions in terms of cost efficiency and complexity lies in the future. For PAM4, we presented a real-time eight-channel DWDM system with already commercially available PHY circuits. In contrast to DMT, PAM4 needs optical dispersion compensation, whereas DMT is able to adapt to the channel characteristics by vestigial side band filtering in combination with bit and power loading, but at the expense of additional OSNR requirement. Bit and power loading also enables seamless adjustment of the data rate without the need of variable clock sources, as shown in the experiments for a four to eight-channel VSB-DMT transmission system.

## Acknowledgements

The results were obtained in the framework of the SASER-ADVAntage-NET and SpeeD projects, partly funded by the German ministry of education and research (BMBF) under contracts 16BP12400 and 13N13744, and by the European Commission in the Marie Curie project ABACUS.

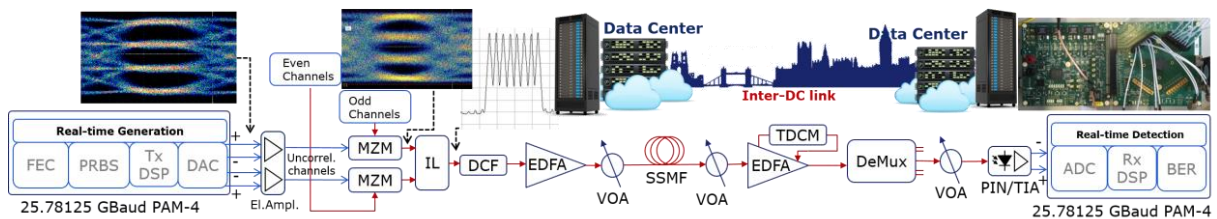

**Fig. 3:** Experimental setup for single channel and DWDM transmission (IL: Interleaver, EDFA: Erbium Doped Fiber Amplifier, TDCM: Tunable Dispersion Compensating Module, VOA: Variable Optical Attenuator).